\documentclass[a4paper,fleqn,usenatbib]{mnras}

\usepackage{mathptmx}

\usepackage[T1]{fontenc}
\usepackage{ae,aecompl}

\usepackage{graphicx}	
\usepackage{amsmath}	
\usepackage{amssymb}	

\newcommand \vhel{\ifmmode{~V_{{\rm HEL}}}\else{~$V_{{\rm HEL}}$}\fi}

\title[The short period binary CSPN of M~3-1]{The short orbital period binary star at the heart of the planetary nebula M~3-1\thanks{Photometric and radial velocity measurements presented in this paper are available at CDS via anonymous ftp to \url{cdsarc.u-strasbg.fr} (\url{130.79.128.5}) or via \url{http://cdsarc.u-strasbg.fr/viz-bin/qcat?J/MNRAS}}}
\author[Jones et al.]{David Jones,$^{1,2}$\thanks{E-mail:
djones@iac.es} Henri~M.~J. Boffin,$^3$, Paulina Sowicka,$^4$ Brent Miszalski,$^{5,6}$ 
\newauthor
Pablo Rodr\'iguez-Gil,$^{1,2}$ Miguel Santander-Garc\'ia,$^7$ and Romano~L.~M. Corradi$^{8,1}$
\\
$^{1}$Instituto de Astrof\'isica de Canarias, E-38205 La Laguna, Tenerife, Spain\\
$^{2}$Departamento de Astrof\'isica, Universidad de La Laguna, E-38206 La Laguna, Tenerife, Spain\\
$^{3}$European Southern Observatory, Karl-Schwarzschild-Str 2, 85748 Garching, Germany\\
$^{4}$Nicolaus Copernicus Astronomical Center, Bartycka 18, PL-00-716 Warsaw, Poland\\
$^{5}$South African Astronomical Observatory, PO Box 9, Observatory 7935, South Africa\\
$^{6}$Southern African Large Telescope Foundation, PO Box 9, Observatory 7935, South Africa\\
$^{7}$Observatorio Astron\'omico Nacional (OAN-IGN), C/ Alfonso XII, 3, 28014 Madrid, Spain\\
$^{8}$GRANTECAN, Cuesta de San Jos\'e s/n, 38712 Breña Baja, La Palma, Spain\\
}

\date{Accepted xxxx xxxxxxxx xx. Received xxxx xxxxxxxx xx; in original form xxxx xxxxxxxx xx}

\pagerange{\pageref{firstpage}--\pageref{lastpage}} \pubyear{2018}

\begin{document}
\label{firstpage}
\pagerange{\pageref{firstpage}--\pageref{lastpage}}
\maketitle

\begin{abstract}
We present the discovery of a 3$h$5$m$ orbital-period binary star at the heart of the planetary nebula M~3-1 - the shortest period photometrically-variable central star known and second only to V458 Vul, in general.  Combined modelling of light and radial velocity curves reveals both components to be close to Roche-lobe-filling, strongly indicating that the central star will rapidly evolve to become a cataclysmic variable, perhaps experiencing a similar evolution to V458 Vul resulting in a nova eruption before the planetary nebula has fully dissipated.  While the short orbital period and near Roche-lobe filling natures of both components make the central binary of M~3-1 an important test case with which to constrain the formation processes of cataclysmic variables, novae and perhaps even supernovae type Ia.
\end{abstract}


\section{Introduction}

Binary central stars are critical for understanding the formation and evolution of planetary nebulae (PNe), being the sole remaining contender for driving the shaping of aspherical and axisymmetric PNe \citep{garcia-segura14,jones17c}.  Furthermore, the short-period (P $\leq$ a few days), post-common-envelope (post-CE) systems have fundamental implications for a wide variety of other astrophysical phenomena, in particular for our understanding of the CE phase itself \citep{demarco08,demarco11b,tocknell14}.

Here, we present the discovery of a close-binary central star in the planetary nebula M~3-1 - one of the shortest orbital period binary central stars known \citep[second only to V458 Vul;][]{rodriguez10short} and the shortest to have been discovered photometrically\footnote{V458 Vul shows irregular short timescale variability consistent with quasi-periodic oscillations but no evidence of modulation with the orbital period \citep{bouzid11}.}.

\begin{figure*}
\centering
\includegraphics[width=0.9\textwidth]{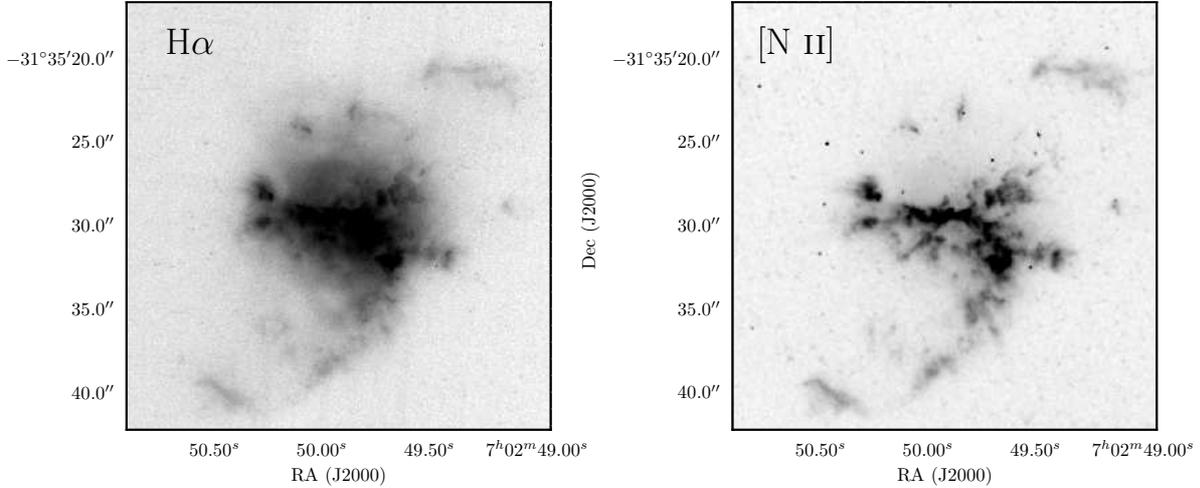}
\caption[]{HST archival imagery of M~3-1 showing its remarkable filamentary waist and extended jet-like structures.}
\label{fig:M3-1ims}
\end{figure*}

M~3-1 ($\alpha$=07$^h$02$^m$49.9$^s$, $\delta$=$-$31$^\circ$35\arcmin{}29.45\arcsec{}, PN~G242.6$-$11.6) has previously been singled out for its remarkable pair of low-ionisation polar ansae \citep{corradi96} which bear a remarkable resemblance to the precessing jets of another planetary nebula known to host a binary central star - Fleming 1 \citep{boffin12b}.  Furthermore, \citet{corradi96} also identified filamentary low-ionisation structures in the equatorial direction, which were argued by \citet{goncalves01} to be fossil remnants of condensations in the old AGB wind.  Higher resolution imagery with the Hubble Space Telescope (HST) reveals further disrupted low-ionisation filaments across the entire central nebula (see figure \ref{fig:M3-1ims}). It is important to note that such features are not only typical of central star binarity  \citep{miszalski09b}, but that a binary is thought to be a necessity in order for their formation \citep{soker94,jones17c}.  Furthermore, these filamentary and jet-like structures are present around several other central star systems which present large-amplitude photometric variability and orbital periods shorter than one day - most notably in Hen~2-11 \citep{jones14a}, Hen~2-155 \citep{jones15}, NGC~6326 and NGC~6778 \citep{miszalski11b}.  As such, M~3-1 was included in our long-term monitoring campaign in search of close-binary central stars through photometric variability.

This paper is organised as follows.  The observations and data reduction are presented in Section \ref{sec:obs}, the modelling of the central star system can be found in Section \ref{sec:mod}, while the results are discussed in Section \ref{sec:conc}.

\section{Observations and data reduction}
\label{sec:obs}
\subsection{Photometry}
\label{sec:phot}
Between February 2012 and March 2015, imaging observations of the central star of M~3-1 were acquired using the EFOSC2 instrument \citep{EFOSC2a,EFOSC2b} mounted on the European Southern Observatory's 3.6-m New Technology Telescope (ESO-NTT).  The E2V CCD (pixel scale 0.24\arcsec{} pixel$^{-1}$) was employed along with the H$\beta$-continuum filter \citep[\#743; as previously employed in][in order to minimise the contamination from the bright nebula]{jones15}. In total 171 images with 150-s exposure time were acquired spread fairly evenly across five observing runs.

The images were debiased and flat-fielded using routines from the \textsc{AstroPy} affiliated package \textsc{ccdproc}.  Differential photometry of the central stars was then performed against field stars using the \textsc{sep} implementation of the \textsc{SExtractor} algorithms \citep{sep,bertin96}, before being placed on an approximate apparent magnitude scale using observations of standard stars taken during the course of the observations.

\begin{figure*}
\centering
\includegraphics[width=\textwidth]{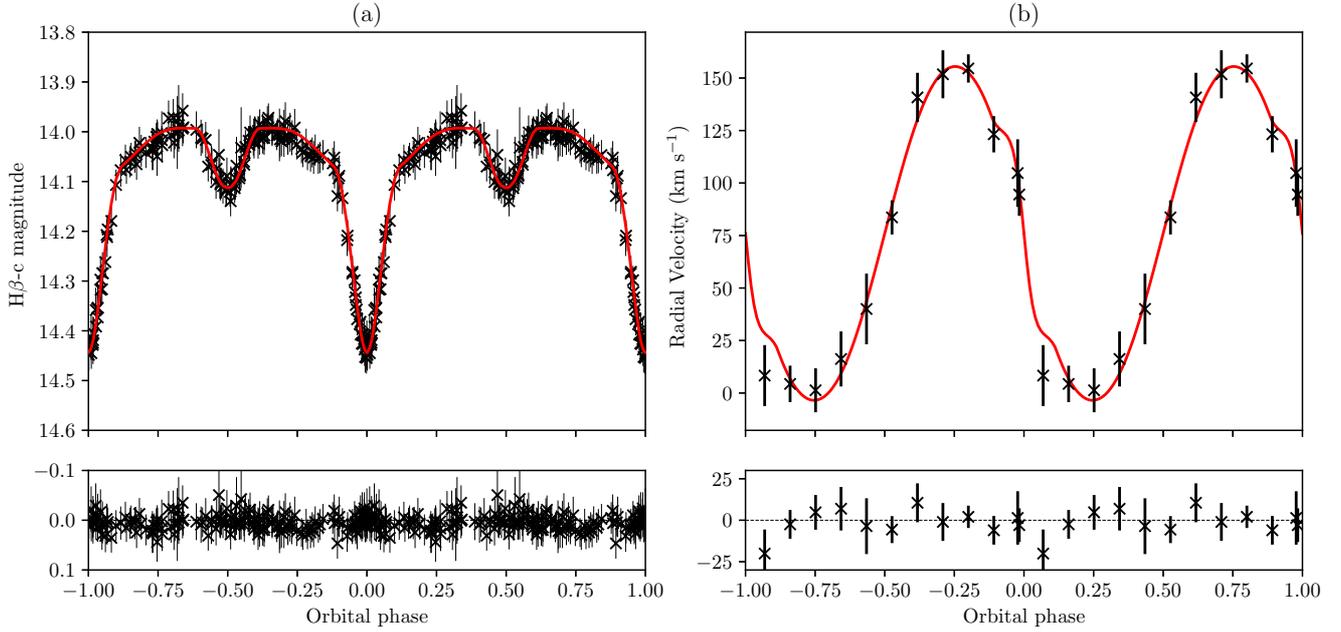}
\caption[]{(a) Phase-folded H$\beta$-continuum light curve of the central star system of M~3-1 (upper panel), as well as the residuals to the modelled curves (lower panel). (b) Phase-folded radial velocity curve of the hot component of the central star system of M~3-1.  The upper panel shows the radial velocity data along with the corresponding curve of the best-fitting model.  The lower panel shows the residuals between the model and the radial velocity data.  Note the presence of deviations from a perfect sine in the model curve at phase $\sim$0 are due to the Rossiter-McLaughlin effect during primary eclipse.}
\label{fig:M3-1all}
\end{figure*}

The observations show clear short-period variability with prominent primary and secondary eclipses (see figure \ref{fig:M3-1all}(a)) which, given the excellent temporal coverage of the data, allow the period to be determined extremely precisely.  The resulting ephemeris is,
\begin{equation}
\mathrm{HJD}=2\,455\,986.6680 (0.0001) +  0.1270971(0.0000001) E
\end{equation}
for the primary eclipse.

\subsection{Radial velocity monitoring}
\begin{figure}
\centering
\includegraphics[width=\columnwidth]{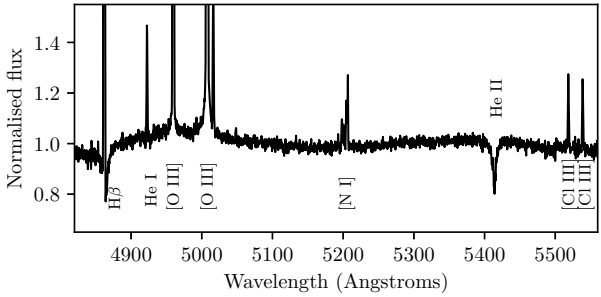}
\caption[]{An example FORS2 spectrum of the central star and nebula of M~3-1, showing the bright nebular lines of [O~\textsc{iii}] and H$\beta$ as well as the clear He~\textsc{ii} stellar absorption line.}
\label{fig:M3-1spec}
\end{figure}

\begin{figure}
\centering
\includegraphics[width=\columnwidth]{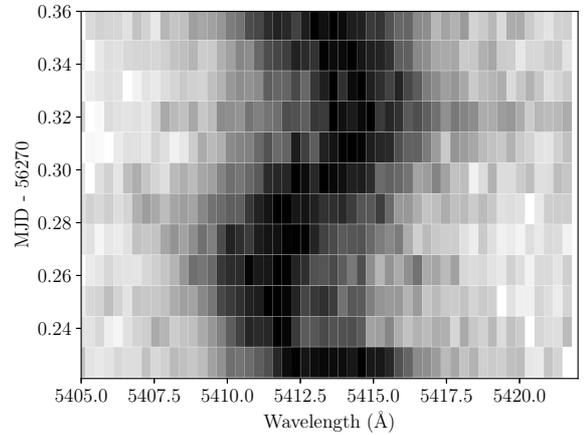}
\caption[]{Trailed FORS2 spectra of the central star of M~3-1 around the He~\textsc{ii} 5412 \AA{} line.}
\label{fig:M3-1trail}
\end{figure}

Spectroscopic monitoring of the central star system of M~3-1 was performed on December 9 2011 using the FOcal Reducer/low dispersion Spectrograph \citep[FORS2;][]{FORSshort} instrument mounted on the Unit Telescope 1 (UT1 or Antu) of the European Southern Observatory's 8.2m Very Large Telescope (ESO-VLT).  A 0.5\arcsec{} wide longslit was employed along with the GRIS\_1400V grism (approximate wavelength coverage, 4600\AA{}$\leq\lambda\leq$5800\AA{}) and the MIT/LL mosaic detector.  Twelve contiguous exposures of 900-s were taken, covering just over one complete orbital period.  Basic reduction (debiasing, flat-fielding, wavelength calibration) was performed using the FORS2 pipeline, before sky-subtraction and optimal extraction of target spectra by standard \textsc{starlink} routines.  The resulting spectra show strong nebular contamination from the bright lines of [O~\textsc{iii}] as well as H$\beta$ (see Figure \ref{fig:M3-1spec}), however the stellar absorption line of He~\textsc{ii} at 5411.52 \AA{} was sufficiently unaffected to allow radial velocities to be measured.  Visual inspection of this line shows clear evidence of sinusoidal variation on timescales consistent with the orbital period derived from the photometry mentioned above (see Figure \ref{fig:M3-1trail}).  No spectral signatures \citep[for example, irradiated emission lines;][]{miszalski11b,jones15} originating from a secondary component were detected in the spectra, in spite of a reasonably prominent irradiation effect in the observed lightcurve. This means that only the radial velocity curve of one component could be derived which, given the spectral signature, is most likely due to the hot nebular progenitor. The final heliocentric radial velocity measurements, as derived via cross-correlation with a template (created from the average profile) of the 5411.52 \AA{} absorption line, are shown phased on the photometric ephemeris in figure \ref{fig:M3-1all}(b).

\section{Modelling}
\label{sec:mod}
The light and radial velocity curves presented were modelled simultaneously using the next-generation Wilson-Devinney code \textsc{phoebe2} \citep{phoebe}.  A fitting was performed via a Markov chain Monte Carlo (MCMC) method implemented in python using {\tt emcee} \citep{emcee} and parallelized to run on the LaPalma supercomputer using {\tt schwimmbad} \citep{schwimmbad}.  Limb-darkening values for the hot, primary component were extrapolated from the tables of \citet{gianninas13}, while the limb-darkening parameters of the secondary were derived using \textsc{phoebe2}.  Initially, the masses, radii and temperatures of both stars were allowed to vary freely.  However, given that the radial velocity curve is single-lined (only the primary's velocities are measured), the results were found to be particularly insensitive to the mass of the primary - with almost any reasonable value (0.4--1.4 M$_\odot$) seemingly similarly likely (though a small, but not statistically significant peak, was found around 0.9 M$_\odot$).  In order to try to break this degeneracy, a canonical value of 0.65 M$_\odot$ was adopted and a second MCMC chain ran.  The final fit parameters are listed in table \ref{tab:M3-1params}, while the resulting light and radial velocity curves are shown overlaid on the observations in figures \ref{fig:M3-1all}(a) and (b), respectively.  It is worth noting that the final values for other parameters are relatively unaffected by the choice of fixed primary mass, with similar values (within a few uncertainties) found for all other parameters (more discussion of the results is presented in the supporting material).

The light and radial velocity curves are well fit with little evidence of systematic deviations between the model and observations.  The radial velocity curve does perhaps show some signs of periodicity in the residuals (lower panel of figure \ref{fig:M3-1all}(b)), however the statistical significance of any variability is rather small given that all points bar one lie within one uncertainty of the model radial velocity curve.  Furthermore, the implied periodicity would be shorter than the measured orbital period and, thus, implausibly short to be due to perturbations arising from the orbit of a third component in the system.  The systemic heliocentric velocity of the system is found to be $\sim$76 km s$^{-1}$, in excellent agreement with the radial velocity of the nebula (70$\pm$15 km s$^{-1}$ determined by  \citealt{schneider83}).  While there are no independent measurements of the nebular inclination (by spatio-kinematic modelling, for example), the images presented in figure \ref{fig:M3-1ims} would seem to indicate that the symmetry axis of the nebula is slightly inclined with respect to the plane of the sky (i.e. not parallel but within perhaps 10--20$^\circ$).  As such, it would appear that the derived binary inclination is consistent with the theoretically predicted alignment between nebula and binary orbital plane \citep{hillwig16}.

The stellar radii are determined with good precision, with only a small dependency on the system's inclination.  Both stars are found to be close to Roche lobe filling, with the primary much larger than a typical evolved white dwarf.  The secondary mass is fairly well-constrained at approximately 0.2 M$_\odot$ - this is a consequence of the star being so close to Roche lobe filling, with greater masses resulting implying too small a radius.  However, it should be noted that this parameter is the most affected by the choice of fixed primary mass and, as such, the quoted uncertainties are almost certainly a lower limit.  In many cases, the main sequence secondaries of post-CE binary central stars have been found to display larger radii than isolated main sequence stars of the same mass \citep{jones15} - in M~3-1 the secondary's mass and radius are both found to be consistent with a late M-type main sequence star.  Perhaps this is a consequence of the close orbit, whereby the secondary cannot be inflated without overflowing its Roche lobe.  The fitting does indicate that the temperature of the secondary is larger than would be expected (at $\sim$7500 K with very large uncertainties), but this is an exceptionally poorly-constrained parameter given the single, narrow-band light curve and the fact that the depth of the secondary eclipse is primarily dominated by the level of irradiation from the primary rather than the secondary's underlying temperature.  Such increased temperatures are similarly observed in many post-CE central star secondaries, but, given the extremely large uncertainties from our fitting, we cannot state conclusively that this is the case in M~3-1.  The temperature of the primary, however, is slightly better constrained (though still with rather large uncertainties) and found to be in excellent agreement with the Hydrogen and Helium Zanstra temperatures derived by \citet{kaler91} at 48 and 65 kK, respectively.  The model primary does not lie particularly close to any of the standard post-AGB evolutionary tracks \citep[e.g.][]{vassiliadis94,bloecker95,millerbertolami16}, however the large uncertainty in temperature translates to a much larger uncertainty on the luminosity meaning that consistent tracks are found from all authors with post-AGB ages less than $\sim$30\,000 years (i.e. the observable lifetime of a PN).  The lack of evolutionary tracks which reproduce the derived values (ignoring uncertainties) is unsurprising given that these tracks are calculated for isolated stars and do not account for the effects of the CE phase or mass transfer \citep{millerbertolami17}.

\begin{table}
\caption{Table of model parameters for the binary central star of M~3-1.  Values marked with an asterisk were fixed in the MCMC fitting, please see text for further details.}             
\label{tab:M3-1params} 
\centering
\begin{tabular}{r c c}
\hline
&Primary & Secondary\\
\hline
T$_\mathrm{eff}$ (K) & 48\,000$^{+17\,000}_{-10\,000}$  & 5\,000--12\,000 \\
Radius (R$_\odot$) & 0.41$\pm0.02$ & 0.23$\pm0.02$ \\
Mass (M$_\odot$) & 0.65* & 0.17$\pm0.02$ \\
\hline
Orbital period (days) & \multicolumn{2}{c}{0.1270971$\pm$0.0000001}\\
Inclination & \multicolumn{2}{c}{75.5$^\circ$$^{+1.9^\circ}_{-1.5^\circ}$}\\
$\gamma$ (km s$^{-1}$)& \multicolumn{2}{c}{76.0$\pm$0.3}\\
$K_1$ (km s$^{-1}$)& \multicolumn{2}{c}{79.5$\pm$4.2}\\
\hline
\end{tabular}
\end{table}

\section{Discussion}
\label{sec:conc}

We have demonstrated the central star of the PN M~3-1 to be a close-binary system with a period of approximately 3$h$5$m$, making it the shortest-period, photometrically-variable binary central star known to-date and the second shortest-period in general \citep[second only to V458 Vul;][]{rodriguez10short}.  Intriguingly, both components of the binary system are found to be very close to Roche-lobe-filling - making the system interesting from the point-of-view of a possible merger candidate.  However, given the low masses of both components, the time to merger via gravitational waves would be approximately 1.5~Gyr.  The primary would almost certainly be expected to settle to a smaller radius (i.e. no longer be Roche-lobe-filling) on a much shorter timescale, meaning that mass transfer will again become feasible.  As such, the orbital separation of the system will evolve as a function of mass exchange, mass loss and magnetic braking rather than the radiation of gravitational waves.

Given the good agreement between the most-likely modelled temperature and the literature temperatures derived by studying the ionisation of the nebula, it is likely that the true central star does not lie on standard post-AGB evolutionary tracks.  This is not to be unexpected given that such tracks are for single stars, and do not account for the poorly-understood CE, but it is an important point to emphasise given that models of post-CE central stars are frequently compared to such tracks.

The secondary of the system is found to be close to Roche-lobe filling but apparently not ``inflated'' with respect to the expected radius given its mass - such inflation is a frequently observed trait in post-CE central stars and is generally thought to be due to accretion prior to or during the CE phase \citep{miszalski13b,jones15,chamandy18}.  As such, it is unlikely that the secondary will thermally adjust to adopt a smaller radius in the near-future, meaning that small changes in the orbital separation could lead to Roche-lobe overflow and accretion onto the hot primary - turning the system into a cataclysmic variable (CV).  This likely evolution draws strong parallels with the only binary central star known to present a shorter orbital period - V458 Vul - a bona fide CV which experienced a nova eruption inside a previously ejected PN \citep{wesson08short}.  The age difference between the nova eruption and the ejection of the planetary nebula (believed to have been formed by the ejection of a CE, just as would be the case for the nebula of M~3-1) is only $\sim$14\,000 years.  While the total mass of the components of M~3-1 is rather uncertain, it is almost certainly lower than that of V458 Vul, and as such it does not represent a possible supernova Ia progenitor system. It does, however, offer a valuable system with which to study the formation and evolution of such supernova progenitors at an intermediate stage of their evolution, when they are ``fresh out of the oven'' of the CE phase, as well as the formation of CVs in general.  The presence of the ejected envelope, as well as the evident jets, offer a unique window with which to study the CE phase which is key in the formation of such objects.  This, coupled with the observation that a significant number of supernovae Ia are observed to explode in circumstellar environments consistent with a previous PN phase \citep{tsebrenko15}, makes M~3-1 (and other short-period binary central stars) important tools with which to understand the supernova Ia phenomenon.  We therefore particularly encourage further study of the nebula, both chemical \citep[e.g.;][]{corradi15,jones16} and morpho-kinematical \citep[][]{corradi11,boffin12b,tocknell14}, as well as further observations of the central star system (particularly multi-colour photometry to better constrain the parameters of both stars) - all of which can reveal important information about the formation and evolution of the system.

\section*{Acknowledgments}
The authors wish to thank the anonymous referee for their useful comments.  This paper is based on observations made with ESO Telescopes at the La Silla Paranal Observatory under programme IDs 088.D-0573, 090.D-0435, 090.D-0693, 091.D-0475, 092.D-0449, 094.D-0091.  This research has been supported by the Spanish Ministry of Economy and Competitiveness (MINECO) under the grant AYA2017-83383-P.  P.S. thanks the Polish National Center for Science (NCN) for support through grant 2015/18/A/ST9/00578. B.M. acknowledges support from the National Research Foundation (NRF) of South Africa. The authors thankfully acknowledge the technical expertise and assistance provided by the Spanish Supercomputing Network (Red Espa\~nola de Supercomputaci\'on), as well as the computer resources used: the LaPalma Supercomputer, located at the Instituto de Astrofs\'ica de Canarias.

\bibliographystyle{mnras}
\bibliography{literature.bib}

\begin{thebibliography}{}
\makeatletter
\relax
\def\mn@urlcharsother{\let\do\@makeother \do\$\do\&\do\#\do\^\do\_\do\%\do\~}
\def\mn@doi{\begingroup\mn@urlcharsother \@ifnextchar [ {\mn@doi@}
  {\mn@doi@[]}}
\def\mn@doi@[#1]#2{\def\@tempa{#1}\ifx\@tempa\@empty \href
  {http://dx.doi.org/#2} {doi:#2}\else \href {http://dx.doi.org/#2} {#1}\fi
  \endgroup}
\def\mn@eprint#1#2{\mn@eprint@#1:#2::\@nil}
\def\mn@eprint@arXiv#1{\href {http://arxiv.org/abs/#1} {{\tt arXiv:#1}}}
\def\mn@eprint@dblp#1{\href {http://dblp.uni-trier.de/rec/bibtex/#1.xml}
  {dblp:#1}}
\def\mn@eprint@#1:#2:#3:#4\@nil{\def\@tempa {#1}\def\@tempb {#2}\def\@tempc
  {#3}\ifx \@tempc \@empty \let \@tempc \@tempb \let \@tempb \@tempa \fi \ifx
  \@tempb \@empty \def\@tempb {arXiv}\fi \@ifundefined
  {mn@eprint@\@tempb}{\@tempb:\@tempc}{\expandafter \expandafter \csname
  mn@eprint@\@tempb\endcsname \expandafter{\@tempc}}}

\bibitem[\protect\citeauthoryear{{Appenzeller} et~al.,}{{Appenzeller}
  et~al.}{1998}]{FORSshort}
{Appenzeller} I.,  et~al., 1998, The Messenger, \href
  {http://adsabs.harvard.edu/abs/1998Msngr..94....1A} {94, 1}

\bibitem[\protect\citeauthoryear{Barbary}{Barbary}{2016}]{sep}
Barbary K.,  2016, \mn@doi [The Journal of Open Source Software]
  {10.21105/joss.00058}, 1

\bibitem[\protect\citeauthoryear{{Bertin} \& {Arnouts}}{{Bertin} \&
  {Arnouts}}{1996}]{bertin96}
{Bertin} E.,  {Arnouts} S.,  1996, A\&AS, \href
  {http://adsabs.harvard.edu/abs/1996A%26AS..117..393B} {117, 393}

\bibitem[\protect\citeauthoryear{{Bloecker}}{{Bloecker}}{1995}]{bloecker95}
{Bloecker} T.,  1995, A\&A, \href
  {http://adsabs.harvard.edu/abs/1995A%26A...299..755B} {299, 755}

\bibitem[\protect\citeauthoryear{{Boffin}, {Miszalski}, {Rauch}, {Jones},
  {Corradi}, {Napiwotzki}, {Day-Jones}  \& {K{\"o}ppen}}{{Boffin}
  et~al.}{2012}]{boffin12b}
{Boffin} H.~M.~J.,  {Miszalski} B.,  {Rauch} T.,  {Jones} D.,  {Corradi}
  R.~L.~M.,  {Napiwotzki} R.,  {Day-Jones} A.~C.,   {K{\"o}ppen} J.,  2012,
  \mn@doi [Science] {10.1126/science.1225386}, \href
  {http://adsabs.harvard.edu/abs/2012Sci...338..773B} {338, 773}

\bibitem[\protect\citeauthoryear{{Bouzid} \& {Garnavich}}{{Bouzid} \&
  {Garnavich}}{2011}]{bouzid11}
{Bouzid} S.,  {Garnavich} P.,  2011, in American Astronomical Society Meeting
  Abstracts \#217. p. 338.10

\bibitem[\protect\citeauthoryear{{Buzzoni} et~al.,}{{Buzzoni}
  et~al.}{1984}]{EFOSC2a}
{Buzzoni} B.,  et~al., 1984, The Messenger, \href
  {http://esoads.eso.org/abs/1984Msngr..38....9B} {38, 9}

\bibitem[\protect\citeauthoryear{{Chamandy} et~al.,}{{Chamandy}
  et~al.}{2018}]{chamandy18}
{Chamandy} L.,  et~al., 2018, preprint, \href
  {http://adsabs.harvard.edu/abs/2018arXiv180503607C} {} (\mn@eprint {arXiv}
  {1805.03607})

\bibitem[\protect\citeauthoryear{{Corradi}, {Manso}, {Mampaso}  \&
  {Schwarz}}{{Corradi} et~al.}{1996}]{corradi96}
{Corradi} R.~L.~M.,  {Manso} R.,  {Mampaso} A.,   {Schwarz} H.~E.,  1996, A\&A,
  \href {https://ui.adsabs.harvard.edu/#abs/1996A&A...313..913C} {313, 913}

\bibitem[\protect\citeauthoryear{{Corradi} et~al.,}{{Corradi}
  et~al.}{2011}]{corradi11}
{Corradi} R.~L.~M.,  et~al., 2011, MNRAS, 410, 1349

\bibitem[\protect\citeauthoryear{{Corradi}, {Garc{\'{\i}}a-Rojas}, {Jones}  \&
  {Rodr{\'{\i}}guez-Gil}}{{Corradi} et~al.}{2015}]{corradi15}
{Corradi} R.~L.~M.,  {Garc{\'{\i}}a-Rojas} J.,  {Jones} D.,
  {Rodr{\'{\i}}guez-Gil} P.,  2015, \mn@doi [ApJ] {10.1088/0004-637X/803/2/99},
  \href {http://adsabs.harvard.edu/abs/2015ApJ...803...99C} {803, 99}

\bibitem[\protect\citeauthoryear{{De Marco}, {Hillwig}  \& {Smith}}{{De Marco}
  et~al.}{2008}]{demarco08}
{De Marco} O.,  {Hillwig} T.~C.,   {Smith} A.~J.,  2008, \mn@doi [AJ]
  {10.1088/0004-6256/136/1/323}, \href
  {http://adsabs.harvard.edu/abs/2008AJ....136..323D} {136, 323}

\bibitem[\protect\citeauthoryear{{De Marco}, {Passy}, {Moe}, {Herwig}, {Mac
  Low}  \& {Paxton}}{{De Marco} et~al.}{2011}]{demarco11b}
{De Marco} O.,  {Passy} J.-C.,  {Moe} M.,  {Herwig} F.,  {Mac Low} M.-M.,
  {Paxton} B.,  2011, \mn@doi [\mnras] {10.1111/j.1365-2966.2010.17891.x},
  \href {http://adsabs.harvard.edu/abs/2011MNRAS.411.2277D} {411, 2277}

\bibitem[\protect\citeauthoryear{{Foreman-Mackey}, {Hogg}, {Lang}  \&
  {Goodman}}{{Foreman-Mackey} et~al.}{2013}]{emcee}
{Foreman-Mackey} D.,  {Hogg} D.~W.,  {Lang} D.,   {Goodman} J.,  2013, \mn@doi
  [PASP] {10.1086/670067}, \href
  {http://adsabs.harvard.edu/abs/2013PASP..125..306F} {125, 306}

\bibitem[\protect\citeauthoryear{{Garc{\'{\i}}a-Segura}, {Villaver}, {Langer},
  {Yoon}  \& {Manchado}}{{Garc{\'{\i}}a-Segura} et~al.}{2014}]{garcia-segura14}
{Garc{\'{\i}}a-Segura} G.,  {Villaver} E.,  {Langer} N.,  {Yoon} S.-C.,
  {Manchado} A.,  2014, \mn@doi [ApJ] {10.1088/0004-637X/783/2/74}, \href
  {http://adsabs.harvard.edu/abs/2014ApJ...783...74G} {783, 74}

\bibitem[\protect\citeauthoryear{{Gianninas}, {Strickland}, {Kilic}  \&
  {Bergeron}}{{Gianninas} et~al.}{2013}]{gianninas13}
{Gianninas} A.,  {Strickland} B.~D.,  {Kilic} M.,   {Bergeron} P.,  2013,
  \mn@doi [ApJ] {10.1088/0004-637X/766/1/3}, \href
  {http://adsabs.harvard.edu/abs/2013ApJ...766....3G} {766, 3}

\bibitem[\protect\citeauthoryear{{Gon{\c{c}}alves}, {Corradi}  \&
  {Mampaso}}{{Gon{\c{c}}alves} et~al.}{2001}]{goncalves01}
{Gon{\c{c}}alves} D.~R.,  {Corradi} R. L.~M.,   {Mampaso} A.,  2001, \mn@doi
  [ApJ] {10.1086/318364}, \href
  {https://ui.adsabs.harvard.edu/#abs/2001ApJ...547..302G} {547, 302}

\bibitem[\protect\citeauthoryear{{Hillwig}, {Jones}, {De Marco}, {Bond},
  {Margheim}  \& {Frew}}{{Hillwig} et~al.}{2016}]{hillwig16}
{Hillwig} T.~C.,  {Jones} D.,  {De Marco} O.,  {Bond} H.~E.,  {Margheim} S.,
  {Frew} D.,  2016, \mn@doi [ApJ] {10.3847/0004-637X/832/2/125}, \href
  {http://adsabs.harvard.edu/abs/2016ApJ...832..125H} {832, 125}

\bibitem[\protect\citeauthoryear{{Jones} \& {Boffin}}{{Jones} \&
  {Boffin}}{2017}]{jones17c}
{Jones} D.,  {Boffin} H.~M.~J.,  2017, \mn@doi [Nature Astronomy]
  {10.1038/s41550-017-0117}, \href
  {http://adsabs.harvard.edu/abs/2017NatAs...1E.117J} {1, 0117}

\bibitem[\protect\citeauthoryear{{Jones}, {Boffin}, {Miszalski}, {Wesson},
  {Corradi}  \& {Tyndall}}{{Jones} et~al.}{2014}]{jones14a}
{Jones} D.,  {Boffin} H.~M.~J.,  {Miszalski} B.,  {Wesson} R.,  {Corradi}
  R.~L.~M.,   {Tyndall} A.~A.,  2014, \mn@doi [A\&A]
  {10.1051/0004-6361/201322797}, \href
  {http://adsabs.harvard.edu/abs/2014A%26A...562A..89J} {562, A89}

\bibitem[\protect\citeauthoryear{{Jones}, {Boffin}, {Rodr{\'{\i}}guez-Gil},
  {Wesson}, {Corradi}, {Miszalski}  \& {Mohamed}}{{Jones}
  et~al.}{2015}]{jones15}
{Jones} D.,  {Boffin} H.~M.~J.,  {Rodr{\'{\i}}guez-Gil} P.,  {Wesson} R.,
  {Corradi} R.~L.~M.,  {Miszalski} B.,   {Mohamed} S.,  2015, \mn@doi [A\&A]
  {10.1051/0004-6361/201425454}, \href
  {http://adsabs.harvard.edu/abs/2015A%26A...580A..19J} {580, A19}

\bibitem[\protect\citeauthoryear{{Jones}, {Wesson}, {Garc{\'{\i}}a-Rojas},
  {Corradi}  \& {Boffin}}{{Jones} et~al.}{2016}]{jones16}
{Jones} D.,  {Wesson} R.,  {Garc{\'{\i}}a-Rojas} J.,  {Corradi} R.~L.~M.,
  {Boffin} H.~M.~J.,  2016, \mn@doi [MNRAS] {10.1093/mnras/stv2519}, \href
  {http://adsabs.harvard.edu/abs/2016MNRAS.455.3263J} {455, 3263}

\bibitem[\protect\citeauthoryear{{Kaler} \& {Jacoby}}{{Kaler} \&
  {Jacoby}}{1991}]{kaler91}
{Kaler} J.~B.,  {Jacoby} G.~H.,  1991, \mn@doi [ApJ] {10.1086/169967}, \href
  {http://adsabs.harvard.edu/abs/1991ApJ...372..215K} {372, 215}

\bibitem[\protect\citeauthoryear{{Miller Bertolami}}{{Miller
  Bertolami}}{2016}]{millerbertolami16}
{Miller Bertolami} M.~M.,  2016, \mn@doi [A\&A] {10.1051/0004-6361/201526577},
  \href {http://adsabs.harvard.edu/abs/2016A%26A...588A..25M} {588, A25}

\bibitem[\protect\citeauthoryear{{Miller Bertolami}}{{Miller
  Bertolami}}{2017}]{millerbertolami17}
{Miller Bertolami} M.~M.,  2017, in {Liu} X.,  {Stanghellini} L.,   {Karakas}
  A.,  eds,  IAU Symposium Vol. 323, Planetary Nebulae: Multi-Wavelength Probes
  of Stellar and Galactic Evolution. pp 179--183 (\mn@eprint {arXiv}
  {1611.09801}), \mn@doi{10.1017/S1743921317001533}

\bibitem[\protect\citeauthoryear{{Miszalski}, {Acker}, {Parker}  \&
  {Moffat}}{{Miszalski} et~al.}{2009}]{miszalski09b}
{Miszalski} B.,  {Acker} A.,  {Parker} Q.~A.,   {Moffat} A.~F.~J.,  2009,
  \mn@doi [A\&A] {10.1051/0004-6361/200912176}, \href
  {http://adsabs.harvard.edu/abs/2009A%26A...505..249M} {505, 249}

\bibitem[\protect\citeauthoryear{{Miszalski}, {Jones}, {Rodr{\'{\i}}guez-Gil},
  {Boffin}, {Corradi}  \& {Santander-Garc{\'{\i}}a}}{{Miszalski}
  et~al.}{2011}]{miszalski11b}
{Miszalski} B.,  {Jones} D.,  {Rodr{\'{\i}}guez-Gil} P.,  {Boffin} H.~M.~J.,
  {Corradi} R.~L.~M.,   {Santander-Garc{\'{\i}}a} M.,  2011, \mn@doi [A\&A]
  {10.1051/0004-6361/201117084}, \href
  {http://adsabs.harvard.edu/abs/2011A%26A...531A.158M} {531, A158}

\bibitem[\protect\citeauthoryear{{Miszalski}, {Boffin}  \&
  {Corradi}}{{Miszalski} et~al.}{2013}]{miszalski13b}
{Miszalski} B.,  {Boffin} H.~M.~J.,   {Corradi} R.~L.~M.,  2013, \mn@doi
  [MNRAS] {10.1093/mnrasl/sls011}, \href
  {http://adsabs.harvard.edu/abs/2013MNRAS.428L..39M} {428, L39}

\bibitem[\protect\citeauthoryear{{Price-Whelan} \&
  {Foreman-Mackey}}{{Price-Whelan} \& {Foreman-Mackey}}{2017}]{schwimmbad}
{Price-Whelan} A.~M.,  {Foreman-Mackey} D.,  2017, \mn@doi [The Journal of Open
  Source Software] {10.21105/joss.00357}, \href
  {http://adsabs.harvard.edu/abs/2017JOSS....2..357P} {2, 357}

\bibitem[\protect\citeauthoryear{{Pr{\v s}a} et~al.,}{{Pr{\v s}a}
  et~al.}{2016}]{phoebe}
{Pr{\v s}a} A.,  et~al., 2016, \mn@doi [ApJS] {10.3847/1538-4365/227/2/29},
  \href {http://adsabs.harvard.edu/abs/2016ApJS..227...29P} {227, 29}

\bibitem[\protect\citeauthoryear{{Rodr{\'{\i}}guez-Gil}
  et~al.,}{{Rodr{\'{\i}}guez-Gil} et~al.}{2010}]{rodriguez10short}
{Rodr{\'{\i}}guez-Gil} P.,  et~al., 2010, \mn@doi [MNRAS]
  {10.1111/j.1745-3933.2010.00895.x}, \href
  {http://adsabs.harvard.edu/abs/2010MNRAS.407L..21R} {407, L21}

\bibitem[\protect\citeauthoryear{{Schneider}, {Terzian}, {Purgathofer}  \&
  {Perinotto}}{{Schneider} et~al.}{1983}]{schneider83}
{Schneider} S.~E.,  {Terzian} Y.,  {Purgathofer} A.,   {Perinotto} M.,  1983,
  \mn@doi [ApJS] {10.1086/190874}, \href
  {http://adsabs.harvard.edu/abs/1983ApJS...52..399S} {52, 399}

\bibitem[\protect\citeauthoryear{{Snodgrass}, {Saviane}, {Monaco}  \&
  {Sinclaire}}{{Snodgrass} et~al.}{2008}]{EFOSC2b}
{Snodgrass} C.,  {Saviane} I.,  {Monaco} L.,   {Sinclaire} P.,  2008, The
  Messenger, \href {http://adsabs.harvard.edu/abs/2008Msngr.132...18S} {132,
  18}

\bibitem[\protect\citeauthoryear{{Soker} \& {Livio}}{{Soker} \&
  {Livio}}{1994}]{soker94}
{Soker} N.,  {Livio} M.,  1994, \mn@doi [ApJ] {10.1086/173639}, \href
  {http://adsabs.harvard.edu/abs/1994ApJ...421..219S} {421, 219}

\bibitem[\protect\citeauthoryear{{Tocknell}, {De Marco}  \&
  {Wardle}}{{Tocknell} et~al.}{2014}]{tocknell14}
{Tocknell} J.,  {De Marco} O.,   {Wardle} M.,  2014, \mn@doi [MNRAS]
  {10.1093/mnras/stu079}, \href
  {http://adsabs.harvard.edu/abs/2014MNRAS.439.2014T} {439, 2014}

\bibitem[\protect\citeauthoryear{{Tsebrenko} \& {Soker}}{{Tsebrenko} \&
  {Soker}}{2015}]{tsebrenko15}
{Tsebrenko} D.,  {Soker} N.,  2015, \mn@doi [MNRAS] {10.1093/mnras/stu2567},
  \href {http://adsabs.harvard.edu/abs/2015MNRAS.447.2568T} {447, 2568}

\bibitem[\protect\citeauthoryear{{Vassiliadis} \& {Wood}}{{Vassiliadis} \&
  {Wood}}{1994}]{vassiliadis94}
{Vassiliadis} E.,  {Wood} P.~R.,  1994, \mn@doi [ApJS] {10.1086/191962}, \href
  {http://adsabs.harvard.edu/abs/1994ApJS...92..125V} {92, 125}

\bibitem[\protect\citeauthoryear{{Wesson} et~al.,}{{Wesson}
  et~al.}{2008}]{wesson08short}
{Wesson} R.,  et~al., 2008, \mn@doi [ApJL] {10.1086/594366}, \href
  {http://adsabs.harvard.edu/abs/2008ApJ...688L..21W} {688, L21}

\makeatother
\end{thebibliography}

\newpage

\appendix
\section{Supporting information: MCMC corner plot}
\label{MCMC}

\begin{figure*}
\centering
\includegraphics[width=\textwidth]{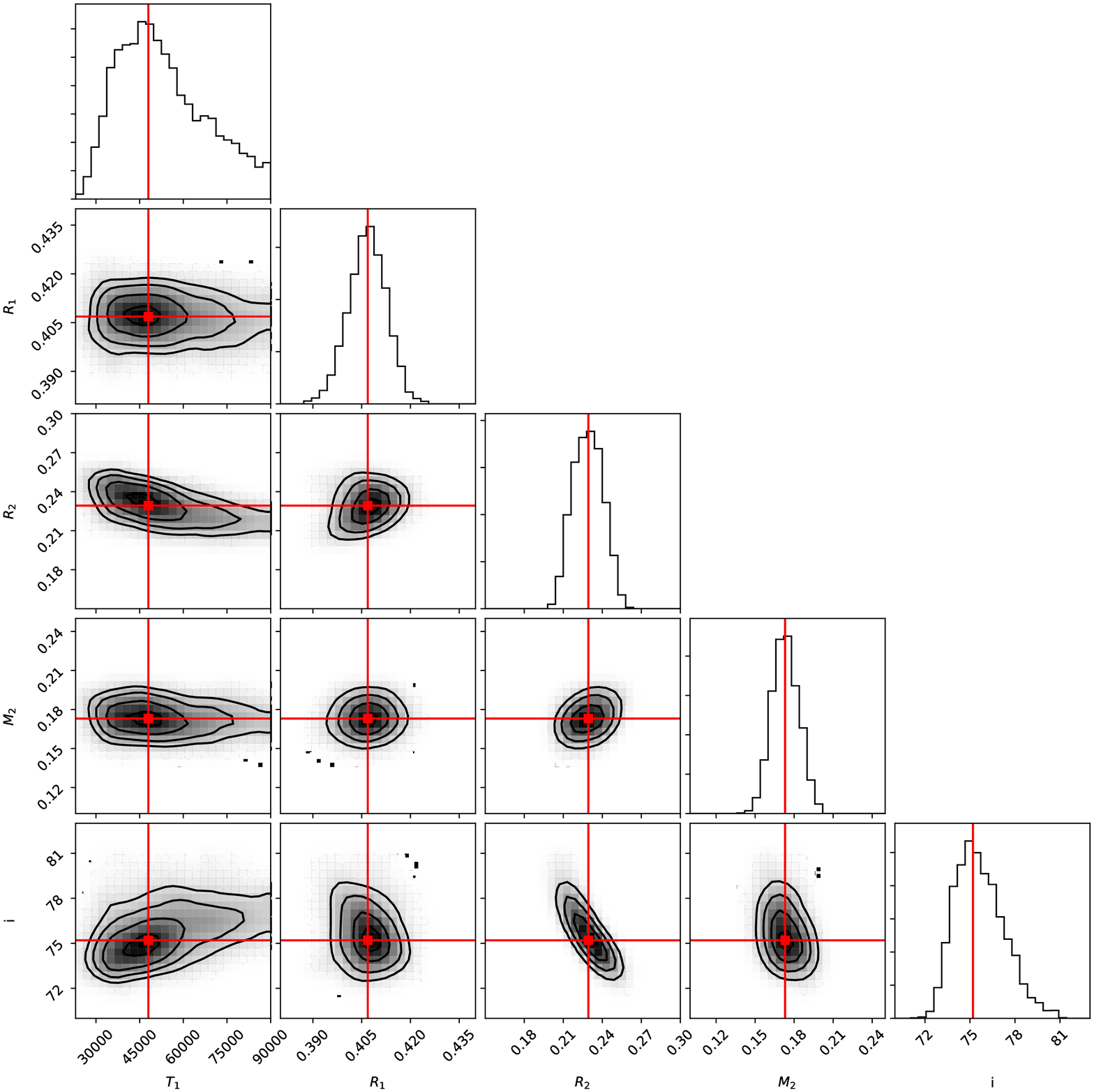}
\caption[]{Corner plot showing the MCMC results for the well-constrained parameters - primary temperature (T$_1$) and radius (R$_1$), secondary radius (R$_2$) and mass (M$_2$), and orbital inclination (i).  Red lines mark the most likely values (as quoted in Table 1).}
\label{fig:M3-1corner}
\end{figure*}

Given that the quoted uncertainties in Table 1 of the main paper do not completely reveal the inter-dependencies between modelled parameters, we present a corner plot of the MCMC chain in Figure \ref{fig:M3-1corner}.  The plot clearly shows the relatively strong anti-correlation between secondary radius and inclination, with most other parameters presenting more gaussian distributions.  The primary temperature is clearly the most poorly constrained of the parameters shown (to be expected given the single- narrow-band nature of the light curve), but does show a clear peak at $\sim$ 50 kK which is consistent with estimates of the temperature based on nebular ionisation.  Furthermore, the primary temperature does not show particularly strong correlations against any other parameters with only a weak correlation with orbital inclination (and weak anti-correlation with secondary radius) - meaning that the uncertainty does not propagate to other parameters, which can be considered more reliably determined.

\label{lastpage}

\end{document}